\documentclass[12pt]{article}
 \usepackage{amsmath,amssymb}
 \newcommand{\be}{\begin{equation}}
 \newcommand{\ee}{\end{equation}}
 \newcommand{\bl}{\begin{equation}\begin{array}{ll}}
 \newcommand{\el}{\end{array}\end{equation}}
 \newcommand{\bll}{\begin{equation}\begin{array}{lll}}
 \newcommand{\bdm}{\begin{displaymath}}
 \newcommand{\edm}{\end{displaymath}}
 \def\bea{\begin{eqnarray}}
 \def\eea{\end{eqnarray}}
 \def\barr{\begin{array}}
 \def\earr{\end{array}}
 \newcommand{\bean}{\begin{eqnarray}}
 \newcommand{\eean}{\end{eqnarray}}
\def\p{\partial}

 \def\La{\Lambda}
 \def\al{\alpha}
 \def\ga{\gamma}
 \def\om{\omega}
 \def\sig{\sigma}

\def\half{\frac{1}{2}}

\def\2third{\frac{2}{3}}
\def\4third{\frac{4}{3}}
\def\3quart{\frac{3}{4}}

\def\sixth{\frac{1}{6}}

\def\pr{\prime}

\def\bg{\bar{g}}

\def\bv{\bar{v}}
\def\bl{\bar{l}}

\def\brho{\bar{\rho}}

\def\bk{\bar{k}}
\def\bl{\bar{l}}

\def\bchi{\bar{\chi}}

\def\cH{{\cal H}}
\def\cL{{\cal L}}

\def\ddrho{\ddot{\rho}}
\def\drho{\dot{\rho}}

\def\dsig{\dot{\sigma}}
\def\dpsi{\dot{\psi}}
\def\dal{\dot{\alpha}}

\def\dbe{\dot{\beta}}

\def\dga{\dot{\gamma}}

\def\dA{\dot{A}}
\def\ddA{\ddot{A}}

\def\dH{\dot{H}}

\def\hep{\hat{\epsilon}}

\def\hr{\hat{r}}

\def\Lag{{\cal L}}

\usepackage{amsmath}

\begin{document}
\raggedbottom

 \title{{\bf A fresh view of cosmological models describing very early Universe: general solution of the dynamical equations.
 }}

 \author{A.T.~Filippov \thanks{Alexandre.Filippov@jinr.ru} \\
{\small \it {$^+$ Joint Institute for Nuclear Research, Dubna, Moscow
 Region RU-141980} }}

 \maketitle

 \begin{abstract}
  The dynamics of any spherical cosmology with a scalar field (`scalaron') coupling to gravity is described by  the nonlinear second-order differential equations for two metric functions and the scalaron depending on the `time' parameter.  The equations depend on the scalaron potential and on the \emph{arbitrary gauge function} that describes time parameterizations. This dynamical system can be integrated for flat, isotropic models  with very special potentials. But, somewhat unexpectedly, replacing the time  variable $t$ by one of the metric functions allows us to \emph{completely integrate} the general spherical theory \emph{in any gauge and with apparently arbitrary potentials}. The main restrictions on the potential arise from positivity of the derived analytic expressions for the solutions, which are essentially the squared canonical momenta. An interesting consequence is emerging of \emph{classically forbidden regions} for these analytic solutions. It is also shown that in this rather general model the \emph{inflationary solutions} can be identified, explicitly derived, and compared to the standard approximate expressions. This approach can be applied to intrinsically anisotropic models with a massive vector field (`vecton') as well as to some non-inflationary models.

 \end{abstract}
  \bigskip \bigskip

 \textbf{1.~Introduction.}
  \smallskip

 In this paper we extend the new method for constructing the exact solutions of FLRW cosmological models with a scalaron (inflaton), which was proposed in \cite{ATFnew}, to  fairly general homogeneous non-isotropic cosmologies with scalaron or vecton. The next step must be to identify, classify, and study different classes of solutions in connection with inflation and other possible cosmological scenarios.  The third step should be to study their interrelations in the frame of the two-dimensional reduction of spherical gravity coupled to scalarons. This program had been attempted to partly implement in \cite{ATF1}-\cite{ATF14}, in the context of the multi-Liouville and Toda-Liouville integrable models, where the three sectors of the solutions were derived and classified. The difficult problem that was not really touched in those papers is the derivation and study of the solutions in presence of small perturbations. Here we discuss only \emph{the first step -- formulating the new approach to constructing exact solutions of a rather general cosmological dynamics with arbitrary potentials $\bv(\al)\equiv v[\psi(\al)]$.}

 We derive the cosmological models by the standard spherical reduction of the four-dimensional theory either of a \emph{scalar field} $\psi$ (scalaron) or a \emph{massive vector field} $A_n$ (vecton) minimally coupled to Einstein's gravity. The scalaron model is characterized by an arbitrary potential $v(\psi)$, which we transform to the potential $\bv(\al)$ depending on the metric function. The vecton model is the linearized version of the models discussed in the papers \cite{ATF}-\cite{ATF13}, which in dimension one is reduced to a scalaron model with a non-canonical kinetic term. The cosmological models are derived by a further dimensional reduction of the two-dimensional scalaron or vecton theories. With this aim, we apply a kind of a direct separating the $r$ and $t$ variables that does not require any group-theoretical considerations (see, e.g., discussion in \cite{VDATF3}-\cite{VDATF2}). Thus we get nonlinear dynamical systems of functions that depend on $t$ or on $r$ and can  describe cosmological models or static states.

  By solving these systems we are finding some \emph{possible} cosmologies (or static black holes) but this does not permit  us to uncover the real processes of how these objects emerge from the solutions of the two-dimensional dynamical system. This is a separate problem not discussed in this paper.\footnote{The author supposes that the reader is familiar with basics of cosmology including modern inflationary models as described in, e.g., \cite{Mukhanov}-\cite{Martin}. In addition, the present paper is not completely independent of \cite{ATFnew}, and some familiarity with it will be helpful (this paper also contains many references to related work). } The stability of various solutions with respect to small higher-dimensional gravitational perturbations was studied in \cite{LifKh}. If the solution is stable, we can regard it as a sort  of attractor in the space of solutions. In this sense, the Friedmann-type solutions must be attractors.

 In next Section, we outline some general reductions to spherical cosmology with the scalaron or vecton. In Sec.~3 we introduce the Lagrangian and Hamiltonian equations in a most general gauge (parametrization of the `time' variable), and discuss different approximations and gauges. The main results are presented in Sec.~4, where we introduce and solve the general gauge independent equations for the scalaron cosmology, which include the anisotropy ($\sig$-function) and curvature ($k$-parameter) dependence. Their explicit analytic solution is possible either in $\sig \equiv 0$ or $k=0$ approximations.\footnote{Here we do not discuss the evident but somewhat cumbersome iterative corrections.} The most important for inflationary models is our exact expression (\ref{mom14}) for $\chi^2(\rho)$ that in particular allows us to generalize the inflationary perturbative expansion of $\chi^2$ in terms of derivatives of $\,l(\psi)\equiv\ln v(\psi)$, \cite{ATFnew}.

 The last Section~5, that was added to this essentially edited and extended version of the paper, discusses the `newborn' problem of positivity of our solutions.

  \bigskip\bigskip
 \textbf{2.~Spherical non-isotropic gravity.}
  \smallskip

 For future considerations we first write the approximate four-dimensional model of gravity  minimally coupled to
 the scalaron $\psi$ and vecton $A_i$ (considered in \cite{ATFn}, \cite{ATFs}):
 \be
 {\Lag}^{(4)}\, = \,  \sqrt{-g}\, \biggr[R[g] - \kappa \biggr(\half F_{ij} F^{ij} + m^2 A_i A^i + g^{ij} \p_i \psi \, \p_j \psi + v(\psi) \biggl) \biggl]\,,
 \label{L1a}%1
 \ee
 where the arbitrary potential $v(\psi)$ may include the cosmological term $2\Lambda$ and, possibly, the mass term for $\psi$. To find the spherically reduced theory we write the metric in the form
 \be
 ds_4^2 = e^{2\alpha} dr^2 + e^{2\beta} d\Omega^2 (\theta , \phi) -
 e^{2\gamma} dt^2 + 2e^{2\delta} dr dt \, ,
 \label{eq1}%2
 \ee
 where  $\alpha, \beta, \gamma, \delta$ depend on $(t, r)$ and  $d\Omega^2 (\theta , \phi)$ is the metric on the 2-dimensional sphere $S^{(2)}$.

 Then  the two-dimensional reduction of four-dimensional theory (\ref{L1a}) can easily be found (the prime denotes differentiations with respect to $r$ and the dot -- with respect to  $t$)\,\footnote{A detailed description of the reduction procedure was given in our earlier work, e.g., \cite{VDATF3}, \cite{VDATF2}.}:
 \be
 \cL^{(2)} = \cL_{gr}^{(2)} + e^{2\beta} \bigl[e^{-\alpha - \gamma} (\dA_1 - A_0')^2 +\,  e^{\alpha - \gamma} (\dpsi^2 + m^2 A_0^2) - e^{-\alpha + \gamma} (\psi'^2 + m^2 A_1^2) - e^{\alpha +\gamma} v(\psi)  \bigr] \,,
 \label{eq2}%3
 \ee
 where $\psi = \psi(t,r)$, $A_i = A_i(t,r)$, $\dA_1 - A_0' \equiv F_{10}\,$,
 \be
 \cL_{gr}^{(2)} \equiv
 e^{-\alpha + 2\beta +\gamma} (2\beta'^2 + 4\beta' \gamma') -
 e^{\alpha + 2\beta - \gamma} (2\dbe^2 + 4\dbe \dal) +
  2\bk e^{\alpha + \gamma}\,,
 \label{eq3}%4
 \ee
  and we omit the total derivatives that do not affect the equations of motion. Notice that we also omit  the terms depending on $\delta$ by taking the formal limit $\delta \rightarrow -\infty$. The variation in this parameter gives the \emph{zero momentum constraint}, which in this limit is simply
 \be
 -{\dot{\beta}}^{\prime} - \dot{\beta} {\beta}^{\prime} +
 \dot{\alpha} {\beta}^{\prime} + \dot{\beta} {\gamma}^{\prime} \,\,
  = \,\, \half \,(\dot{\psi} {\psi}^{\prime} + A_0 A_1) \,.
 \label{A2}%5
 \ee
 All other equations are obtained by varying the metric functions $\al, \beta$.

 To find possible cosmological models we require $\psi^\pr \equiv 0$, $A_0 \equiv 0$ ($A_1 \equiv A(t)$) and make further reductions by separating the variables $t$ and $r$ in the metric,
 \be
 \alpha = \alpha_0(t) + \alpha_1(r) , \quad \beta = \beta_0(t) +
 \beta_1(r) , \quad \gamma = \gamma_0(t) + \gamma_1(r) \,.
 \label{A3}%6
 \ee
 The momentum constraint severely restricts possible cosmological models that can be derived from the two-dimensional Lagrangian. It is not difficult to find that  there are four types of cosmological and static solutions.  Two main solutions are: $\beta^\prime = \gamma^\prime=0$ (`\emph{general anisotropic'} cosmology)\footnote{We call the `\emph{special' anisotropic} the cosmology with $\beta^\prime = \gamma^\prime$ and $\dal=0$, which is dual to FRLW. The \emph{flat isotropic} cosmology is obtained from the general anisotropic one if in addition $\bk = k=0$ and $\dot{\alpha}=\dot{\beta}$.} and $\dot{\alpha}=\dot{\beta},\, \gamma^\prime=0$ (general isotropic, or, FLRW-cosmology). For other solutions of (\ref{A2}) there emerge strong restrictions on $v(\psi)$, requiring $v_\psi\equiv0$ or even $v\equiv0$. The two-dimensional theory reduced by conditions $\psi^\prime=0$, (\ref{A3}), and (\ref{A2}) describe  homogeneous cosmologies with scalaron $\psi(t)$ if we require their three-dimensional curvature to be constant (notice unusual definition of $\bk, k$),
 \be
 R^{(3)} \equiv\,g^{ij}\,R^{(3)}_{ij} = 2\bk\,e^{-2 \beta_1} - 2\,e^{-2\al_1}(\,2{\beta_1}'' +\, 3 {\beta_1'}^2 - 2\,\beta^\pr_1\al^\pr_1)\,=\, -6k\,.
 \label{A4}%7
 \ee
 In addition, the three-dimensional subspace is isotropic if
 \be
 R^1_1=-2\,e^{-2\al_1}(\,{\beta_1}'' +\, {\beta_1'}^2-\,\beta^\pr_1\al^\pr_1)\,=\,R^2_2=\bk\,e^{-2 \beta_1} - e^{-2\al_1}({\beta_1}'' +\,2{\beta_1'}^2 - \beta^\pr_1\al^\pr_1)=R^3_3\,,
 \label{A5}%8
 \ee
 which is  equivalent to the relation
 \be
 \bk\,e^{-2(\beta_1-\al_1)} + {\beta_1}'' - \beta^\pr_1\al^\pr_1\,=\,0\,.
 \label{A6}%9
 \ee
 Taking into account that we can choose $\al^\pr=0$, the two conditions are equivalent to one:
 \be
 {\beta_1'}^2 - \bk \,e^{-2 \beta_1} \,=\, 3k \,.
 \label{A7}%10
 \ee

 These relations also directly follow from the conditions $\dot{\alpha}=\dot{\beta},\, \gamma^\prime=0$ and from the requirement of separation in the equation of motion. We discussed these matters in more detail in papers quoted above. Here we only remind that the `general anisotropic'  cosmology (with $\beta^\prime = \gamma^\prime=0$) becomes isotropic if $\bk=0$; then also $k=0$ but nevertheless, for the general solution, $\dal \neq \dbe$ as is demonstrated below. We will immediately see that, in this case, there exists the unique solution with $\dot{\alpha}=\dot{\beta}$. The 4-dimensional solution then necessary satisfies $\gamma^\prime=0$ condition and therefore it is the true FLRW-type solutions.

 From now on we ignore the subtleties related to higher-dimensional interpretation of the solutions and only consider (one-dimensional) cosmological dynamical equations.

  \bigskip \bigskip
 \textbf{3.~General dynamical equations}
  \smallskip

 Now let us introduce cosmological reductions of theories (\ref{L1a}). As was shown in \cite{ATFn}, the approximate cosmological Lagrangian\,\footnote{In the cosmological models of \cite{ATF}-\cite{ATFs} with the massive vecton $A_i$ there is no $\psi$. In the small $\dA_i^2$ approximation there remain from its kinetic part the term $\dA_i^2$ and the constant potential $v=2\La$.   }
 can be written in the form ($A \equiv A_z (t)$):
 \be
 \cL_c =
 e^{2\beta} \bigl[e^{-\alpha - \gamma} \dA^2 -
 e^{-\alpha + \gamma} m^2 A^2 - e^{\alpha +\gamma} v(\psi)  -
 e^{\alpha - \gamma} (2\dbe^2 + 4\dbe \dal - \dpsi^2) \bigr] - 6k\,e^{\al+\ga} \,.
 \label{eq10}%11
 \ee
 To write the equations of motion in a more compact form, we introduce notation
 \be
 3\,\rho \equiv (\alpha + 2\beta) \,, \quad
 3\,\sigma \equiv (\beta-\alpha) \,,\quad \al = \rho-2\sig\,; \quad
 3\,A_{\pm} = e^{-2\rho + 4\sigma} (\dA^2 \pm
 m^2 e^{2\gamma} A^2) \,.
 \label{eq11}%12
 \ee
 Then the exact Lagrangian for the vecton-scalaron cosmology is:
 \be
 \cL_c = e^{3\rho - \gamma} (\dpsi^2 - 6\drho^2 + 6\dsig^2)\,-\,e^{3\rho + \gamma}\, v(\psi) - \,6k\,e^{\rho-2\sig+\ga} \,+\, e^{3\rho\,-\,\gamma}\, 3A_{-}\,.
 \label{eq12}%13
 \ee
 Here $e^\gamma$ is the Lagrange multiplier, variations of which yield the \emph{energy constraint}:
 \be
 \cH_c \equiv \dpsi^2 - 6\drho^2 + 6\dsig^2 + e^{2\gamma}\,v(\psi) + 6k\,e^{2\ga-2(\rho + \sigma)} + 3\,A_{+} \,=\,0\,.
 \label{eq13}%14
 \ee
 As in any gauge theory with one constraint of this type,
 we can choose one gauge fixing condition (the standard
 ones are $\gamma = 0$, $\gamma = \alpha$, $\gamma = 3\alpha$). The other equations are
 \be
 4\,\ddot{\rho} + 6\,\drho^2 -4\,\drho \dga  + 6\,\dsig^2
 + \dpsi^2 - e^{2 \gamma} \, v(\psi) \,=\,2k\,e^{2\ga-2(\rho + \sigma)}\,-\,A_{-} \,,
  \label{eq14}%15
 \ee
 \be
 \ddot{\sigma}\,+\,(3\,\drho\,-\,\dga)\,\dsig\,=\,k\,e^{2\ga-2(\rho + \sigma)}\,+\,A_{-} \,,
  \label{eq16}%16
 \ee
 \be
 \ddot{\psi} + (3\,\drho - \dga)\,\dpsi \,+\,e^{2 \gamma} \, v^\pr(\psi)/2 = 0;
 \label{eq14a}%17
 \ee
 \be
 \ddA + (\drho + 4\dsig - \dga) \dA + e^{2\gamma}\,m^2 A = 0 \,,
 \label{eq14b}%18
 \ee
 Equation (\ref{eq14}) can be replaced by two equations obtained by subtracting from it Eq.(\ref{eq13}),
 \be
 \ddot{\rho} + (3\,\drho - \dga)\,\drho - e^{2\gamma}\,v(\psi)/2 \,=\,2k\,e^{2\ga-2(\rho + \sigma)} \,+\,(3 A_+\,-\,A_-)/4 \,,
 \label{eq15}%19
 \ee
 or alternatively, by adding them together (note that $\,3A_+ \pm A_-\geq 0\,$ if $\,m^2>0$),
 \be
 \ddot{\rho}\,+\,3\,\dsig^2 \,-\,\drho \dga\,+\,\dpsi^2/2 \,=\,-k\,e^{2\ga - 2(\rho+\sigma)}\,-\,(3 A_+\,+\,A_-)/4\,.
  \label{eq15a}%20
 \ee

 First, a few general remarks on equations (\ref{eq13})-(\ref{eq15a}). This system can be simplified by applying a \emph{gauge fixing condition} while some subsystems can be solved in any gauge. Further, taking account of the fact that the sums of the first two terms in equations (\ref{eq16})-(\ref{eq15}) are essentially time derivatives of the  momenta of the dynamical variables $(\rho, \psi, \sig)$ and $A$, \be
 (p_\rho,\,p_\psi,\,p_\sig) = 2\,e^{3\rho-\ga}(-6\drho,\,\dpsi, \,6\,\dsig)\,,\quad p_A = 2\,e^{\rho + 4\sig -\ga} \dA\,,
 \label{mom1}%21
 \ee
 we see that these definitions together with equations (\ref{eq16})-(\ref{eq15}) are the Hamilton equations with the canonical Hamiltonian (compare to the pure scalaron model in \cite{ATFnew}):
 \be
 \cH_c^{\textrm{can}} = \frac{1}{24}(6p_\psi^2 + p_\sig^2 - p_\rho^2 + 6p_A^2\,e^{2\rho-4\sig})\,e^{\ga-3\rho} \,+\,
 v(\psi)\,e^{\ga+3\rho} \,+\, 6k\,e^{\ga+\rho-2\sig} \,+\, m^2 A^2\,e^{\ga+\rho+4\sig}\,.
 \label{mom2}%22
 \ee

 In \cite{ATFnew} we exploited only the simplest gauge choices, which in the present general model look like $\ga + c\rho =0$. When $A\equiv\,0,\,k=0$, the gauge $\ga-3\rho = 0$ is most useful (unlike the `standard' gauge $\ga=0$). More general gauges are $\,g(p,q;\ga)=0$.\footnote{Here $q =\psi, \rho, \sig,  A$ and $p$ -- the corresponding momenta. As $q$ are independent of $\ga$, a simple and safe gauge condition is $\,g\,(q;\ga)=0\,$ with $\,\p_\ga g\neq 0$. The restriction on the general $g$ is $\,\p_p\,g\,\p_\ga p\,+\,\p_\ga g \neq 0$.} Thus if we try to choose a gauge in which r.h.s. of (\ref{eq16}) vanishes we must express $\dA^2$ (in  $A_-$) in terms of $p_A\,$. Then we find that this condition does not fix $\ga$ because all the terms are proportional to $e^{2\ga}$. Other interesting properties of these equations are: independence of the $\psi$-equation of $k, \sig, A$ and independence of the $A$-equation of $k, \psi$. Note that $\drho(t)\equiv \xi(t)$ is the most important cosmological function, which in the isotropic limit coincides with the standard \emph{Hubble parameter} (function)  usually denoted by $H(t)$. We keep the same notation for our generalized `Hubble function' $\drho(t)$. Equations (\ref{eq15})-(\ref{eq15a}) give useful restrictions on $\dH(t)$:
 \be
 \dH(t)\equiv \ddrho(t) \geq 0,\quad \textrm{if}\quad \ga=3\rho,\, v\geq 0,\,\,k,\,\geq 0;\,\qquad \dH(t)\leq 0,\quad \textrm{if}\quad \ga=0,\, k\geq 0.\,
 \label{mom3}%23
 \ee
 By the way, inequality $\dH \geq 0$ for $v\geq 0,\,\,k\,\geq 0$ \emph{keeps in any gauge}. This may hint at idea of the gauge independent solution, our main result presented in next section.

 Formally, there exist six essentially different special cases to consider. Let us denote them by the symbol $\{S, V; C\}$, where:  $S=\psi\,\,\textrm{or}\,\,0$, $V=A\,\,\textrm{or}\,\,0$, $C=k\,\,\textrm{or}\,\,0$. In this paper, we only solve the general case $\{\psi, 0; k\}$ with arbitrary $k$. When $k=0$, the solution coincides with the standard FLRW solution, if we \emph{assume} that $\sigma \equiv 0$. When $k\neq0$, we have not the FLRW cosmology because $\sigma \neq 0$ and, if even $\sigma\rightarrow 0$, the 4-dimensional solution does not coincide with FLRW, for which $\beta^\pr \neq 0$. The essentially different and much more difficult cases $\{0, A;k\}$ and $\{\psi, A;k\}$ will be considered in a separate publication. The interesting and important property of the general system of equations is independence of the $\psi$-equation of $\,k, \sig, A$ and independence of the $A$-equation of $\,k, \psi$.

  \bigskip \bigskip
 \textbf{4.~Gauge independent solution of scalaron cosmology}
  \smallskip

 Now we turn to the  $\{\psi, 0; k\}$ equations which we rewrite in the first-order form for the momentum-like variables $\xi\,,\eta\,, \zeta$ treated as functions of $\rho$\,(using $d/dt = \xi\,d/d\rho$),\,\footnote{In \cite{ATFnew}, we have exactly solved equations for $\xi(\al),\,\eta(\al)$ and approximately -- the corresponding equations for $\xi(\psi),\,\eta(\psi)$. The function  $\chi(\al)\equiv\psi^\pr(\al)$ connects the $\psi$ and $\al$ pictures; here we replace it by $\chi(\rho)\equiv\psi^\pr(\rho)$. }
 \be
 (\drho,\,\dpsi,\,\dsig)\,\equiv [\,\xi(\rho)\,,\eta(\rho)\,,\zeta(\rho)\,] = [\,\xi(\rho),\,\xi\,\psi^\pr(\rho),\,\xi\,\sig^\pr(\rho)\,]\, \equiv \,\xi(\rho)[\,1,\,\chi (\rho),\,\om(\rho)]\,,
 \label{mom4}%24
 \ee
 where $\chi(\rho) \equiv \eta/\xi = \psi^\pr(\rho)$ and $\om(\rho)\equiv \zeta/\xi =\sig^\pr(\rho)$ are gauge invariant functions, the integrals of which, $\psi(\rho)$ and $\sig(\rho)$, give a \emph{portrait} of the scalaron cosmology. If we could derive the portrait, we would find all the characteristics of the cosmology. In paper \cite{ATFnew} we thoroughly studied $\chi(\al)$ and its another incarnation $\bchi(\psi) \equiv d\xi/d\psi = 1/\chi[\,\rho(\psi)]$. When $\sig \equiv 0$ the $\chi$ function can be derived if we take as the input either potential $\bv(\rho)$ or the generalized `Hubble function' $H(\rho) \equiv \xi(\rho)$.\,\footnote{When $k\,\neq 0$, equation (\ref{eq16}) has no solution $\dsig \equiv 0$, but $\dsig$ may be exponentially small if $\rho \rightarrow +\infty$. For this reason, FLRW models with $\sig \equiv 0$ and $k=0$ are meaningful as explained above.} The $\chi$-equation with the $v(\psi)$ input can be exactly solved for a mere handful of potentials. Actually, it was studied in various approximations, described in \cite{ATFnew}, and we here confine ourselves to equations with potentials depending on $\rho$.

 First, let us transform the dynamical equations into first-order form by generalizing approach of \cite{ATFnew}. In $\{\psi,\,0;\,k\}$ case, the dynamical system consists of equations (\ref{eq16})-(\ref{eq14a}) plus (\ref{eq15}) or (\ref{eq15a}).  The constraint (\ref{eq13}), being its integral, is automatically satisfied when the arbitrary integration constant is chosen zero. Then the only problem remaining is the presence of $v^\pr(\psi)$ and of the $\sig$-dependence when $k\neq0$. The formal substitution $\bv(\rho) = v[\psi(\rho)])$ allows us to solve all equations in the $\rho$-version and, as soon as we derive $\chi(\rho)$, we will be able to find the potential in the $\psi$-version, $v(\psi)=\bv[\rho(\psi)]$ for arbitrary $\bv(\rho)$. The obvious relation,
 \be
 v^\pr(\psi) = \frac{dv}{d\psi} = \frac{dv}{d\rho} \frac{d\rho}{d\psi} = \bv^\pr(\rho)\frac{\xi}{\eta} = \bv^\pr(\rho) /\chi(\rho)\,,
 \label{mom5}%25
 \ee
 then solves the $v(\psi)$ problem. Comparing the definitions (\ref{mom4})-(\ref{mom5}) and their discussion with the corresponding consideration in \cite{ATFnew}, it is not difficult to find that (\ref{eq16}),(\ref{eq14a}),(\ref{eq15}) with $A_\pm=0$, $k=0$ are linear differential equations for $\xi^2,\,\eta^2,\,\zeta^2$ in any gauge $\ga(\rho)$ and can be exactly solved for any potential $\bv(\rho)$. The $k$-terms are proportional to $\exp{[-2\sig(\rho)]}$ but, in cosmological consideration, we may suppose that $\sig \ll \rho$ as will be argued below.

 Rewriting the dynamical equations in terms of the \emph{positive gauge-invariant functions},
 \be
 \textbf{S}(\rho)\equiv[\,x(\rho),\,y(\rho),\,z(\rho)\,] \equiv \,\exp{(6\rho-2\ga)}\, [\,\xi^2(\rho)\,,\eta^2(\rho),\,\zeta^2(\rho)\,]\,,
 \label{mom6}%26
 \ee
 we see that (\ref{eq14a}) gives the only equation independent on the curvature term $\sim ke^{4\rho -2\sig}$:
 \be
 y^\pr(\rho)+V^\pr(\rho) - 6V(\rho) = 0\,, \qquad V\equiv e^{6\rho}\,\bv(\rho)\,.
 \label{mom7}%27
 \ee
 Two other equations depend on the curvature term and thus the whole system is not closed:
 \be
 x^\pr(\rho) - V(\rho) = 4k\,e^{4\rho -2\sig},\,\,\,\, \textrm{(a)} \qquad z^\pr(\rho) = 2k\,e^{4\rho -2\sig}\,\sig^\pr(\rho).\,\,\,\, \textrm{(b)}
 \label{mom8}%28
 \ee
 The additional equation that makes it closed is the definition of $\zeta(\rho)$ written as $(\sig^\pr)^2 = z/x \equiv \om^2(\rho)$. This will allow us to show that $\sig$ satisfies a closed equation.

 We first find the formal solution of equations (\ref{mom7})-(\ref{mom8})  plus constraint (\ref{eq13}), which is
 \be
 6\,x(\rho)\,=\,y(\rho) + V(\rho)\,+\,6z(\rho)\,+\,6k\,e^{4\rho-2\sig}\,.
 \label{mom9}%29
 \ee
 To this end, it is sufficient to integrate equations (\ref{mom7}), (\ref{mom8}\,a) and use (\ref{mom9}):
 \be
 y(\rho) = 6 \biggl(C_y + \int V(\rho)\biggr) -  V(\rho)\,, \qquad x(\rho) = \biggl(C_x + \int V(\rho)\biggr) + 4k\int e^{4\rho-2\sig(\rho)}\,,
 \label{mom10}%30
 \ee
 where the integration (in all integrals) is taken over the interval $(-\infty \leq  \rho_0\,,\,\rho)$, while
 \be
 z \equiv x(\rho)\,{\sig^\pr}^2(\rho) = C_z + k\biggl[4\int e^{4\rho-2\sig(\rho)} - e^{4\rho-2\sig(\rho)}\biggl]\,\equiv\, C_x - C_y  + 2k\int \sig^\pr(\rho)\,e^{4\rho-2\sig(\rho)}\,.
 \label{mom11}%31
 \ee
 Eq.(\ref{mom11}) would be nonlinear integro-differential equation for $\sig(\rho)$ if we knew $x(\rho)$.\footnote{It can be transformed into the second-order differential equation for $\sig(\rho)$ but this is not a useful idea.} Supposing $\sig \ll 1$ we can derive $\sig$ by some sort of iterations. The first approximation for $x$, with $\sig=0$, coincides with
 the result of \cite{ATFnew}, because $z\equiv 0$ and therefore $C_x=C_y$. The generalized function $\chi^2$ in this case also coincides with the previous result. The new function, which is a gauge-invariant measure of anisotropy, is given by $\om^2 = z(\rho)/x(\rho)=(\sig^\pr)^2$. From (\ref{mom11}) we see that, when $k=0$, there exists solution $\om^2(\rho)=C_z/x$. If $V(\rho)>0$ and grows with $\rho$, there may exist non-isotropic solutions that become (almost) isotropic for large $\rho$. The isotropic solution $z\equiv 0$ is, probably, the limiting or enveloping one for solutions with $C_z\neq 0$.

 In \emph{summary}, there are four types of the solutions: 1)~ $k=0,\,\sig^\pr=0,\,C_z=0$ -- isotropic solution coinciding with the FLRW cosmology; 2)~$k=0,\,\sig^\pr\neq 0,\, C_z\neq 0$ -- the simplest anisotropic solution, which for a wide class of the potentials $\bv(\rho)$ becomes isotropic at large $\rho$, i.e. $\sig^\pr \rightarrow 0$ for $\rho \rightarrow +\infty$; 3)~$k\neq 0,\,\sig^\pr = 0,\, C_z= 0$ -- isotropic cosmology, not of FLRW type; 4)~$k\neq 0,\,\sig^\pr \neq 0$ -- general anisotropic solution; under some restrictions on $\bv$, the $\sig(\rho)$ can vanish at $\rho \rightarrow +\infty$. We can also show that for a wide enough class of the potentials $\bv(\rho)$ the $\chi(\rho)$-function for large $\rho$ turns out to be small. If we define the generalized conditions for inflation (\emph{ginflation}) by the two inequalities, $\om^2 \ll 1$ (small anisotropy) and $\chi^2 \leq 1$ we find that the generalized \emph{second inflationary parameter} $\hr$ can be defined as follows:\,\footnote{In \cite{ATFnew}, we also define gauge-invariant $\hr(\al)$ but use the definition $y\equiv \eta^2 e^{-2\ga}$ with $\ga=0$. The generalized \emph{first inflationary parameter} is simply $\hep(\rho) \equiv\, \chi^2(\rho)/2$. In folklore on inflation it is approximately constant.}
 \be
 \hr(\rho)\equiv \dpsi^2 e^{-2\ga}/v(\psi)\,=\,y(\rho)\,e^{-6\rho}/\bv(\rho) \,=\,y/V = \chi^2 (1+6k\,e^{-2\rho}/\bv)\,[\,6(1-\om^2)\,-\, \chi^2]^{-1}\,.
 \label{mom12}%32
 \ee
 The r.h.s. coincides with Eq.(75) of \cite{ATFnew} if $\om =0$ and $\rho=\al\,$. Note that, for positive potentials, this relation requires $0<\chi^2<6(1-\om^2)$. Now we can \emph{mutatis mutandis} repeat considerations of paper \cite{ATFnew}, Sec.~4.3, and derive the most important relation:
 \be
 \hr(\rho)\,=\,\frac{6\,C_y}{V(\rho)}\,+\,\frac{6}{V}\int V(\rho)\,-1\,=\,\frac{6\,C_y}{V(\rho)}\,+\,\sum_1^{\infty} (-1)^n\, \frac{\bv^{(n)}(\rho)}{6^n\,\bv(\rho)}\,.%%=\,\frac{\chi^2\, %%(1+6k\,e^{-2\rho}/\bv)}{6\,(1-\om^2) - \chi^2}\,.
 \label{mom13}%33
 \ee
 This relation is the generalization of equation (78) in \cite{ATFnew}, the basic tool for \emph{inflationary perturbation theory} generated by iterations of Eq.(82). The expression of $\hr$ in terms of the functions $\chi^2$ and $\om$ is also valid for arbitrary values of $k$. Eq.(78) is reproduced if we take $\om(\rho)\equiv0,\,k=0$ that corresponds to the above solutions (\ref{mom10}) with $C_x=C_y$ (this is also meaningful when $k\neq0$).  For a wide class of the potentials $\bv(\rho)$ the $C_y$-corrections to $\chi^2$ are exponentially small for large $\rho$. Moreover, supposing  invariance of $\chi^2$ under \emph{scaling of the potential} (see \cite{ATFnew}), $C_y$ and $C_x$ must be zero for the inflationary solution.

 If $k\neq 0$, the $k$-term in (\ref{mom10}) is of order $k\,e^{4\rho}$ for $\rho\rightarrow +\infty$, $\sig(\rho) \rightarrow 0$. Using the expansion in (\ref{mom13}) we can estimate the behavior of $\chi^2$ for large $\rho$ if $\,6\,C_i/V(\rho) \ll\bl^\pr(\rho)< 1$, where $\bl^\pr(\rho)\equiv \bv^\pr(\rho)/\bv(\rho)\equiv
 \chi\,v^\pr(\psi)/v(\psi)$.\,\footnote{Note that inflationary conditions  do not suppose $\rho \rightarrow +\infty\,$, only $\rho \gg 1$. On transition to $v(\psi)$ see \cite{ATFnew}.} Solving Eq.(\ref{mom13}) w.r.t. $\chi^2$, we find the exact formula,
 \be
 \chi^2 = 6(1-\om^2)\biggr[ \sum_1^{\infty} (-1)^n\, \frac{\bv^{(n)}(\rho)} {6^n\,\bv(\rho)}\,+\,\frac{6\,C_y}{V}\, \biggl]  \biggr[1 + \sum_1^{\infty} (-1)^n\, \frac{\bv^{(n)}(\rho)} {6^n\,\bv(\rho)}\,+\,\frac{6}{V}\biggl(k\,e^{4\rho}\,+\,C_y\biggr) \biggl]^{-1}\,,
 \label{mom14}%34
 \ee
 and the first terms of its inflationary expansion (supposing the series converging):
 \be
 \chi^2 = (1-\om^2)\biggr[\biggl(-\bl^{\pr}+\, \textrm{o}(\bl^{\pr})\biggr)+\,36\,C_y\,\frac{e^{-6\rho}}{\bv(\rho)} \biggl] \biggr[\biggr(1 - \frac{1}{6}\,\bl^{\pr}+\, \textrm{o}(\bl^{\pr})\biggr)+\,6\,\frac{e^{-2\rho}}{\bv(\rho)} \biggl(k\,+\, C_y\,e^{-4\rho}\biggr)\biggl]^{-1}\,.
 \label{mom15}%35
 \ee
 This expression also demonstrates that for inflationary solution the finite $k$ corrections may be larger than anisotropic ones. To better understand this, let us look once more on $\om^2(\rho)$. When $C_y/V \equiv C_y\,e^{-6\rho}/\bv(\rho)$ is negligible and $\om^2\ll 1$ we can use the results derived in \cite{ATFnew},\footnote{The inflationary solution corresponds to $C_y=0$, small $C_x=C_z\,$, $\om^2$, and $\chi^2<6$.}  but now we have to add some iteration scheme for deriving $\om^2$, which we cannot discuss here. When $k=0$ we however have the simple gauge-independent expression\footnote{The general solution of Eq.(\ref{eq16}) with $k=A_{-}\equiv0\,$ in the gauge $\ga=3\,\rho\,$ is $\sig=C(t-t_0)$. In any gauge, there exists the integral $\dsig=C\exp(\ga-3\,\rho)$ from which  $\sig(\rho)$ can be derived if we know $\xi(\rho)$.}
 \be
 \om^2(\rho)=\frac{C_z}{x(\rho)}=C_z \biggr[C_x + \int V(\rho) \biggr]^{-1}\,=\,\,6\frac{C_z}{V}\biggl[\biggl(1- \sixth \bl^\pr(\rho) + \textrm{o}(\bl^{\pr})\biggr) +\,6\frac{C_x}{V}\biggl]^{-1}\,.
 \label{mom16}%36
 \ee
 As argued in \cite{ATFnew}, to get a pure inflationary solution we suppose that $C_y=0$, but then, to consider the anisotropic corrections we must take $C_x=C_z\neq0$. In any case, the gauge invariant `anisotropy' remains small and can be neglected in the scalaron inflationary models. The $k\neq 0$ corrections may be more important although they are small for large $\rho$.

 Anyway, fast vanishing of the anisotropic and curvature corrections for large $\rho$ is rather evident property of our solutions for a wide class of the potentials $\bv(\rho)$. This fact sufficiently justifies neglecting these correction in inflationary scenarios.

  \bigskip \bigskip
 \textbf{5.~Remarks on positivity of the solutions}
  \smallskip

 The general solution in the scalaron cosmology, $\textbf{S}(\rho)$, depends on two arbitrary constants $C_y\,$, $C_z\equiv C_x - C_y\,$ and on the parameter $k$. Formally, this solution is defined on  $-\infty < \rho < +\infty$. However, depending on the potential and arbitrary constants (\,including $k$) the dynamical functions may become negative in some domains while they must be positive by definition. Let us introduce some convenient definitions: we say that $\textbf{S}(\rho)$ is positive (negative, zero) at some point $\rho_a$, if all the functions $x,y,z$ are positive (negative, zero) at this point. We also will use the corresponding notation $\textbf{S}(\rho_a)>0$ ($<0,\,=0$). To discuss the physics applications of our solution, we should know the domain ${\cal S}_+$, on which they are positive. Let us consider the class of the potentials, for which ${\cal S}_+$ consists of $N$ intervals, $S_+\equiv \bigcup\,(\,\rho_i, \brho_i)$ (including the possibility of  $\rho_1=-\infty$ and $\brho_N=+\infty$). On the intervals $(\,\brho_i, \rho_{i+1})$ in which $\textbf{S}<0$, the classical solutions do not exist. In other words, these intervals form a sort of `\emph{classically forbidden' regions} and serious discussions of this fact should involve considering elements of quantum cosmology to be attempted in future.

 Here we only remark that the interval $(\rho_0, \rho_1)$, where $0<\rho_1\leq 1$ and $-\infty<\rho_0<0$, the classical picture is certainly not valid. If our classical dynamical system can be used in the interval  $(\rho_1, \rho_+)$, where $\rho_1 \ll  \rho_+$, we may say that there exist two distinct intervals: 1.~`pure classical', $(\rho_1, \rho_+)$, 2.~`simple quantum', $(\rho_0, \rho_1)$, in which a simple quantization approach can be applied (see, e.g., \cite{Hal}-\cite{Lehners}).\,\footnote{For $\rho < \rho_0$ the simple quantum mechanics is inapplicable -- this is the domain of `new physics', see \cite{Linde-mult}.} As a first step one might try to quantize the `truncated' classical system,  $\textbf{s}_k(\rho)\equiv [\,x(\rho), y(\rho)]$, which is isotropic ($\sig \equiv 0$) and becomes `inflationary', when $C_x=C_y=k=0\,$ and $\chi^2<1$. We call this \emph{inflationary Ansatz} (assumption).

 It is clear that even the formal solution of equations (\ref{mom10}) is not completely defined before we find its domain of positive definiteness, where $\textbf{s}_k(\rho)>0$. For any regular potential $V(\rho)$ that exponentially vanishes at $-\infty\,$, an appropriate choice of the integration constants obviously allows to ensure  positivity of $\,\textbf{s}_0(\rho)$ in the domain $D_-\equiv (-\infty, \rho_-)$ with any finite $\rho_-\,$. With such `inflationary' potentials, we can also take the lower integration limit $\rho_0=-\infty$. Now applying the inflationary Ansatz to approximation for $\chi^2$ in Eq.(\ref{mom15}) we find
 \be
 \chi^2 = -\bl^{\pr}(\rho) +\,\textrm{o}(\bl^{\pr}) = -\chi\,v^\pr(\psi)/v(\psi) + ...\,; \quad \chi = -v^\pr(\psi)/v(\psi) +...\equiv -l^\pr(\psi)+\textrm{o}(l^{\pr})\,,
 \label{mom17}%37
 \ee
 which is of course the first term of the inflationary perturbation theory of Ref.~\cite{ATFnew}.\footnote{This simple approximation for inflationary models holds for \emph{inflationary potentials} having $ \rho_-\gg 1$.}

 In general,  supposing that $\bv(\rho)$ and $\sig(\rho)$ are continuous and bounded on any finite interval $\rho_i<\rho<\brho_i$, $i=1,...,N$, we find that there exist the integration constants $C_x\geq0,\,C_y\geq0$, for which the solution is positive on the sum of $N$ such intervals $D_\textrm{\,f}\equiv \bigcup\,(\,\rho_i, \brho_i)$. With additional restrictions on the behavior of the potential at $\rho\rightarrow \pm\infty$  the \emph{support of any positive solution} $\textbf{s}_0(\rho)$ consists of finite sums, ${\cal S}_+ = D_-\bigcup D_\textrm{\,f}\bigcup D_+\,$, where any of the summands may be zero.\footnote{We assume $D_-$ and $D_+\equiv (\rho_+, +\infty)$ simply connected, but $D_\textrm{\,f}\,$ may consist of several intervals.} The complementary set, on which the solution is negative, has the similar structure.

 Potentials that can be useful in cosmological applications are simple enough and the positivity support can often be derived. The simplest example is the exponential potential $V(\rho)=v_0\,e^{\,\bg\rho}$ discussed in \cite{ATFnew}. Especially simple are the inflationary solutions produced by choosing $\rho_0=-\infty$ and $C_x = C_y =0$. Supposing that $\sig(\rho)\equiv 0$, we find the following sufficiently general positivity conditions,
 \be
 6I(\rho) - V(\rho) > 0\,,\qquad I(\rho) + k\,e^{4\rho} >0 \,,\,\,\, \textrm{where} \quad I(\rho) \equiv \int_{-\infty}^{\rho} V(\rho)\,.
 \label{mom18}%38
 \ee
 The second condition can be replaced by the stronger one, $V +\,6k\,e^{4\rho}>0$. For the exponential potential, $\,V = v_0\,e^{\bg\rho}\,$, where $\bg \equiv (g+6)>0$,  the solution (and $\chi^2(\rho)$) is positive everywhere \emph{if and only if}: $v_0>0,\,g<0,\,k\geq 0;\, C_x\geq0,\, C_y\geq0$. For the general multi-exponential potential, $V(\rho)=\,\sum v_i\,e^{\,\bg_i\rho}$,  it is easy to prove that:  when $\,0<\bg_i<6\,$,  $v_i\geq0\,$, $k\geq0$, then conditions (\ref{mom17}) are satisfied and the solution $[\,x(\rho),\, y(\rho)]$ is positive for $-\infty< \rho< +\infty$ provided that $C_x\geq0,\,C_y\geq0$. Relaxing the restrictions on $\,\bg_i<6$, and $v_i\,$, we can derive more complex structures of the positivity domains. The simplest instructive exercise for the reader is analyzing this structures for bi-exponential potentials.

 A very interesting simple potential is $\,V = v_0\,e^{\bg\rho}\,(\rho^2+\,\rho_0^2)$, with $\bg\equiv g+6>0$,
 $v_0>0$ (we suppose that $k=0$ but allow for negative values of $\rho_0^2\,$). It is easy to find that
 $\,\textbf{s}_0(\rho)>0\,$ for all $\,\rho\,$ if and only $\,(g+3)<[\,g\,\bg\,\rho_0]^2/12\,$. Otherwise,
 $S_+ = (-\infty\,,\rho_-) \bigcup\,(\,\rho_+\,, + \infty)$.

 In general case, the functions $\,x(\rho),\,y(\rho)\,$ are positive on systems of disconnected intervals.
 This means that a global structure of the (real) physical solution $\xi(\rho),\, \eta(\rho)$ should be
 better described on the complex $\rho$-plane.

  \bigskip \bigskip
 \textbf{Acknowledgment}

 The author is grateful to J.Halpern for invariable, invaluable  support.

 \end{document}